\documentclass[a4paper,fleqn,usenatbib]{mnras}
\usepackage[T1]{fontenc}
\usepackage{ae,aecompl}
\usepackage{graphicx}	
\usepackage{subfigure}
\usepackage{amsmath}	
\usepackage{amssymb}	
\DeclareMathOperator{\sinc}{sinc}
\defcitealias{MAD12}{Paper I}
\defcitealias{MA14}{Paper II}
\defcitealias{Ray11}{Ray et al. 2011}
\newcommand{\ppcc}{$\,$pc$\,$cm$^{-3}$}	
\newcommand{\tsys}{T$_{\rm sys}$}	
\title[Detection of gamma-ray pulsar J1732$-$3131 at 327\,MHz]
{Detection of radio emission from the gamma-ray pulsar J1732$-$3131 at 327\,MHz}
\author[Yogesh Maan et al.]{Yogesh Maan,$^{1,2}$\thanks{E-mail: maan@astron.nl}
M. A. Krishnakumar,$^{2,3}$ Arun K. Naidu,$^{2}$
\newauthor
Subhashis Roy,$^{2}$ Bhal Chandra Joshi,$^{2}$ Matthew Kerr$^{4}$ and P. K. Manoharan$^{2,3}$ \\
$^{1}$Netherlands Institute for Radio Astronomy (ASTRON), PO Box 2, 7990 AA Dwingeloo, The Netherlands\\
$^{2}$National Centre for Radio Astrophysics, Tata Institute of Fundamental Research, Pune, India\\
$^{3}$Radio Astronomy Centre, NCRA-TIFR, Udagamandalam, India\\
$^{4}$Space Science Division, Naval Research Laboratory, Washington, DC 20375-5352, USA
}

\date{Accepted XXX. Received YYY; in original form ZZZ}
\pubyear{2017}

\begin{document}
\label{firstpage}
\pagerange{\pageref{firstpage}--\pageref{lastpage}}
\maketitle

\begin{abstract}
Although originally discovered as a \emph{radio-quiet} gamma-ray pulsar,
J1732$-$3131 has exhibited intriguing detections at decameter wavelengths.
We report an extensive follow-up of the pulsar at 327\,MHz with the Ooty
radio telescope. Using the previously observed radio characteristics,
and with an effective integration time of 60\,hrs, we present a detection
of the pulsar at a confidence level of 99.82$\%$. The 327\,MHz mean flux
density is estimated to be 0.5$-$0.8\,mJy, which establishes the pulsar
to be a steep spectrum source and one of the least luminous pulsars known
to date. We also phase-aligned the radio and gamma-ray
profiles of the pulsar, and measured the phase-offset between the main peaks
in the two profiles to be 0.24$\pm$0.06. We discuss the observed phase-offset
in the context of various trends exhibited by the radio-loud gamma-ray pulsar
population, and suggest that the gamma-ray emission from J1732$-$3131 is best
explained by outer magnetosphere models. Details of our analysis leading to
the pulsar detection, and measurements of various parameters and their
implications relevant to the pulsar's emission mechanism are presented.
\end{abstract}




\section{Introduction}
The number of gamma-ray pulsars detected by the large area telescope (LAT)
on-board Fermi satellite has gone much beyond the most optimistic
guesses published prior to its launch. This revolution in the population
of gamma-ray pulsars is also well reflected by a very significant
increase in the number of \emph{radio-quiet} gamma-ray pulsars
\citep[35 so far;][]{Caraveo14}. Given the statistically significant numbers
of \emph{radio-quiet} and \emph{radio-loud} pulsars and their
sensitive follow-ups at radio-frequencies, models predicting the
gamma-ray emission regions in the outer magnetosphere are favored
against those supporting the emission sites to be
near the polar cap.
\par
J1732$-$3131 is one of the LAT-discovered pulsars, with a rotation period
of about 196~ms
(see Table~1 for some other parameters of the pulsar).
The early radio searches for its counterpart at high
radio frequency \citep[1374\,MHz;][]{Ray11} turned out to be unsuccessful.
However, searches at decameter wavelengths resulted in an intriguing detection
of a faint signal at the expected period of the pulsar at a dispersion measure
(DM) of 15.44$\pm$0.32\ppcc \citep[][hereafter Paper I]{MAD12}. The pulsar
was detected in only one of the several observing sessions, indicating the
sporadic nature of the radio emission. Detection of several mildly bright
single pulses in the same observing session, at a DM consistent with that
of the periodic signal, further substantiated the findings and suggested
the pulsar to be active at radio frequencies \citepalias{MAD12}. Subsequent
\emph{deep} search at 34\,MHz also resulted in evidences
of very faint periodic signal from the pulsar, and provided more robust
estimate of the flux density \citep[][hereafter Paper II]{MA14}.
\begin{table*}
 \centering
  \caption{Measured and derived parameters of PSR J1732-3131 \citepalias[from][and this work]{Ray11,MAD12,MA14}.}
  \begin{tabular}{lrrlrr}
  \hline
  \hline                                                                
  Right Ascension, R.A.(J2000)          &       &       17:32:33.54     &        Dispersion Measure, DM(pc/cc)       &       &       $15.44\pm0.32$\\
  Declination, Dec. (J2000)             &       &       -31:31:23.0     &        Distance (kpc)                      &       &       $0.60\pm0.15$\\
  Pulse Frequency, $\nu$ ($s^{-1}$)     &       &       5.08794112      &        Flux density at 327\,MHz (mJy)      &       &       0.5$-$0.8\\
  Frequency first derivative, $\dot{\nu}$ ($s^{-2}$)&& -7.2609$\times10^{-13}$&  Spectral index                      &       &       $-3.0$~to~$-2.4$\\
  Epoch of frequency (MJD)              &       &       54933.00        &        pseudo-luminosity at 1400\,MHz ($\mu$Jy\,kpc$^2$)& &     2.2$-$8.9\\ 
\hline
\end{tabular}
\end{table*}
\par
A likely explanation for the apparent lack of radio emission from the
radio-quiet pulsars is that their narrow radio beams miss the sightline
towards Earth \citep{BJ99,WR11}. Since the radio emission beam is
expected to become larger at low frequencies \citep[radius-to-frequency
mapping;][]{Cordes78}, probability of our line-of-sight passing through
the beam also increases. If the above detections of J1732$-$3131 at very
low frequencies were indeed due to this fact, the pulsar can be
expected to be a steep spectrum source. Detections at 34~MHz, when
combined with non-detections at high radio frequencies, imply the spectral
index to be steeper than $-2.3$. Deep observations of the pulsar at
frequencies above 100~MHz could provide a reasonable estimate of the spectral
index, and might even shed some light on its viewing geometry.
\par
Since sky position of J1732$-$3131 is close to the centre of the Galaxy,
the earlier high frequency radio searches would have suffered a loss in
sensitivity due to the enhanced system temperature (\tsys). The upper limit
on 1374~MHz flux density of the pulsar is 59~$\mu$Jy \citep{Ray11}.
For comparison, the L-band flux densities of the LAT-discovered pulsars
J0106+4855 and J1907+0602 are only 8 and 3~$\mu$Jy, respectively
\citep{Pletsch12a,Abdo10}. Hence, flux density of J1732$-$3131 at higher
radio frequencies might still be well within the range of already detected
pulsars, and a deep integration could compensate for the high \tsys~and
help in detection.
\par
Significant variability in the radio flux density of pulsars, at
a range of timescales (from a single rotation period to several hundreds
of seconds, or even larger),
is well known. In the absence of a good understanding of possible
radio emission from the radio-quiet gamma-ray pulsars, their extensive
radio follow-ups might also be revealing. Recent detections of several
energetic radio bursts from the Geminga pulsar \citep{Maan15} demonstrates
the possible rewards of an extensive follow-up. The first radio detection
of J1732$-$3131 \citepalias{MAD12}
suggested the presence of occasional bursty emission from the pulsar.
This provides a strong motivation for a dedicated follow-up.
\par
With the motivations stated above, we conducted deep observations of
J1732$-$3131 using the Ooty radio telescope (ORT). Details
of these observations and data processing are given in Section~2, followed
by a detailed discussion on the results in Section~3. A summary of our
findings is given in Section~4.
\section{Observations and Data Processing}
\subsection{Observations and pre-search processing}
Observations were conducted using the ORT situated
in southern India \citep{Swarup71}. The telescope is equatorially mounted,
and has an offset
parabolic cylindrical reflector with dimensions of 530\,m and 30\,m in
north-south and east-west directions, respectively. With the mechanical
steering in the east-west direction, sources can be tracked continuously
for approximately 9 hours. In the north-south direction, the telescope
beam is steered electronically to point at different declinations (DEC).
\par
Extensive observations of J1732$-$3131 were carried out in March 2014 and
April 2015. In March 2014, the source was observed for a total of 70.5~hours,
distributed over 36 sessions. Typically, 2 observing sessions were conducted
on every day. In April 2015, 37 observing sessions were conducted, amounting
to a total of 54.5~hours of observing time. Each observing session
was accompanied by a few minutes observation of a nearby strong pulsar,
B1749$-$28, which was used as a \emph{control source}.
During each of the individual observing sessions, data were recorded in
filterbank format, with 1024 channels across 16\,MHz bandwidth centered
at 326.5\,MHz, with a time resolution of 0.512\,ms, using the new pulsar
receiver \citep{Naidu15}. For simplicity, we refer the centre frequency
as 327\,MHz in the rest of the paper.
\par
The filterbank data were used to identify the parts of data contaminated
by radio frequency interference (RFI). The identification procedure involves
computation of robust mean and standard deviation, and comparing the data
with a specified threshold separately in the time and frequency domains.
More details of this procedure can be found in \citet{Maan15} and
\citetalias{MA14}. The RFI-contaminated time samples as well as spectral
channels are excluded from any further processing.
\subsection{Search procedures}
The data were searched for transient as well as periodic signals from the
pulsar. Briefly, the search for transient signals involves dedispersing
the filterbank data for a number of optimally spaced trial DMs within
the range 0$-$50\ppcc, computing smoothed versions of each of the dedispersed
timeseries for a number of trial pulse-widths, and searching for events
above a specified threshold.
\par
The filterbank data from individual sessions were searched for any dispersed
signal at the expected rotation period of J1732$-$3131. Since the rotation
ephemeris is known from timing of the gamma-ray data, periodic radio flux
from the pulsar can be probed deeply by combining data from multiple sessions.
Deep search for periodic emission from the pulsar involves folding the
individual session filterbank data over the rotation period, adding the
folded filterbank data from all the sessions in phase, and searching for
a dispersed signal. Detailed descriptions of our single pulse as well as
periodicity searches can be found in \citetalias{MA14} and \citet{ythesis}.
\begin{figure*}
\begin{center}
\includegraphics[width=0.65\textwidth,angle=-90]{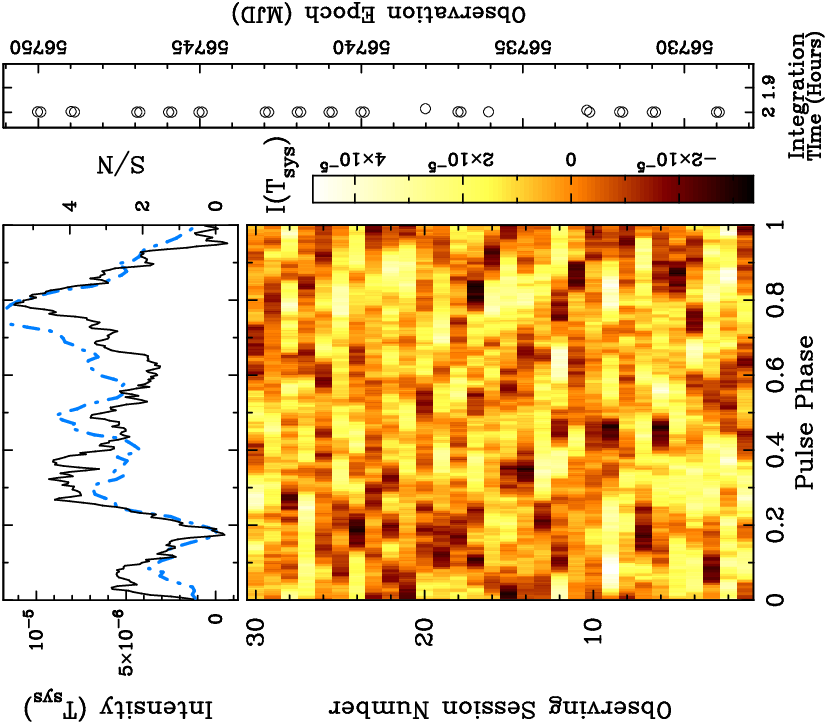}
 \caption{Individual rows in the colour image show phase-aligned 327\,MHz
average profiles of J1732$-$3131 obtained from different sessions (arranged
in ascending order of observing epoch). The upper panel displays the net
average profile (solid line) and the average profile from the earlier
detection at 34\,MHz (dotted line; see Paper I), for ready comparison.
The intensity range of the 34\,MHz profile is normalized to that of the
327\,MHz profile. The 327\,MHz intensity is plotted in units of \tsys,
which is estimated to be 760\,K towards the pulsar (see Section 3.4).
All the individual profiles are smoothed by a 30\degr\,($\sim0.08$ in normalized
pulse-phase)
wide window. For the sessions corresponding to different rows in the image
panel, the right panel shows the epochs of observation and the effective
integration time (earliest epoch corresponds to first observing session).}
 \label{fig_mep1}
\end{center}
\end{figure*}
\begin{figure}
\begin{center}
\includegraphics[width=0.5\textwidth,angle=-90]{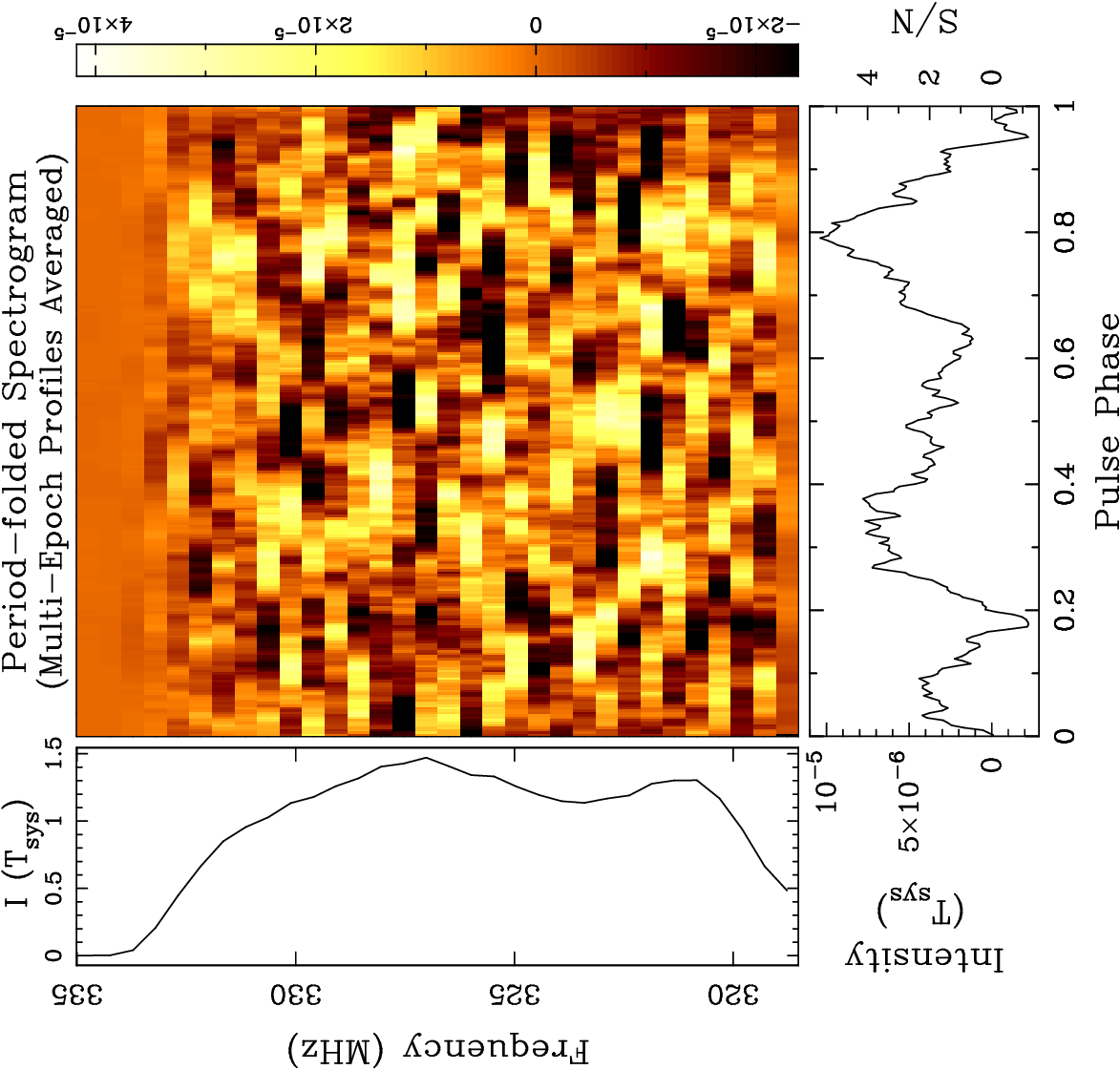}
 \caption{The color image shows dedispersed, period-folded and multi-epoch
averaged 32-channel spectrogram of J1732$-$3131. The left and the bottom
panels show the average spectrum and the net average profile, respectively,
plotted in units of T$_{sys}$. As in Figure~\ref{fig_mep1}, all the profiles
are smoothed by a 30$^{\circ}$ wide window.}
\label{fig_mep2}
\end{center}
\end{figure}
\section{Results and Discussion}
Our searches for bright dispersed pulses and periodic signal using data
from individual observing sessions did not result in any significant detection
above a signal-to-noise ratio (S/N) threshold of 8$\sigma$. More details of
the deep searches using
data from March 2014 and April 2015 are discussed below separately.
\begin{figure}
\begin{center}
\includegraphics[width=0.3\textwidth,angle=-90]{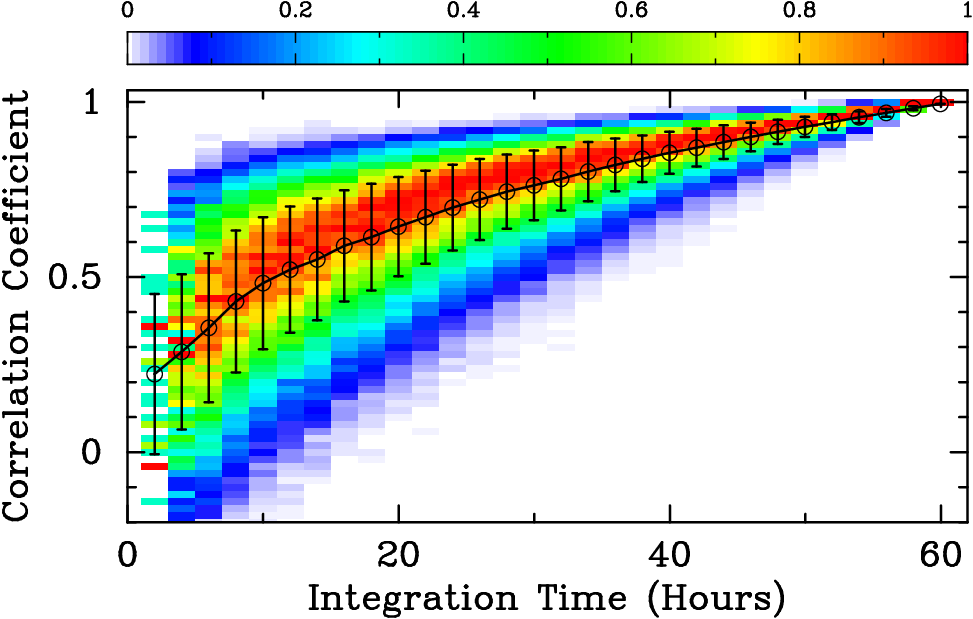}
\caption{Each column in the colour-image shows the distribution function of
\emph{normalized} cross-correlation coefficient between partial average
profiles corresponding to the integration time on horizontal axis and the
net average profile. Peaks of all the distribution functions are normalized
to 1. The `open circles' connected by the continuous black line show the median
correlation coefficient for each of the trial integration time. The errorbars
on either side of the median correspond to the standard deviation measured
from the distribution function.}
\label{fig_boot}
\end{center}
\end{figure}
\begin{figure*}
\centering
\subfigure{\includegraphics[width=0.2\textwidth,angle=-90]{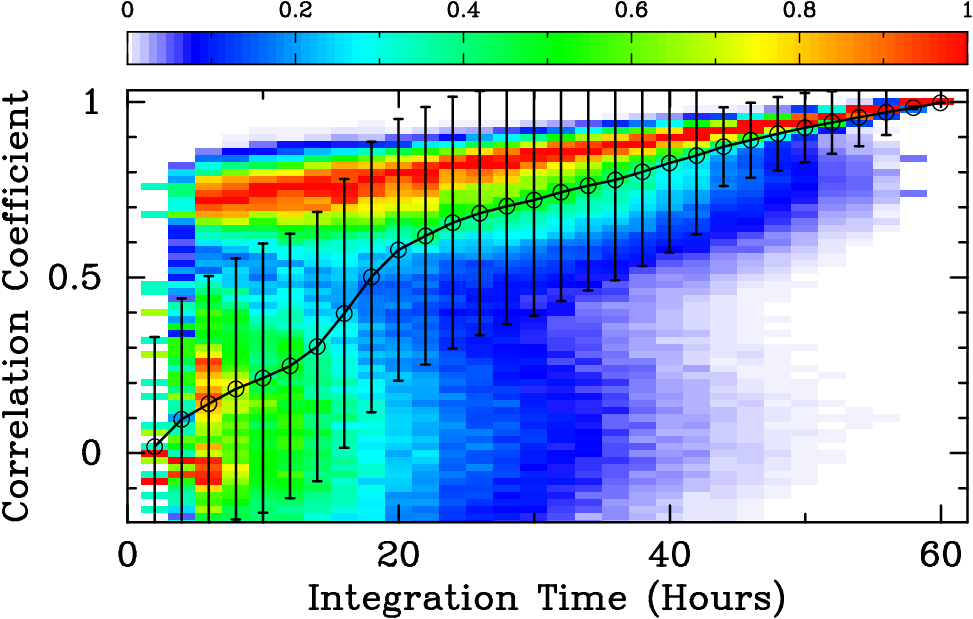}}
\subfigure{\includegraphics[width=0.2\textwidth,angle=-90]{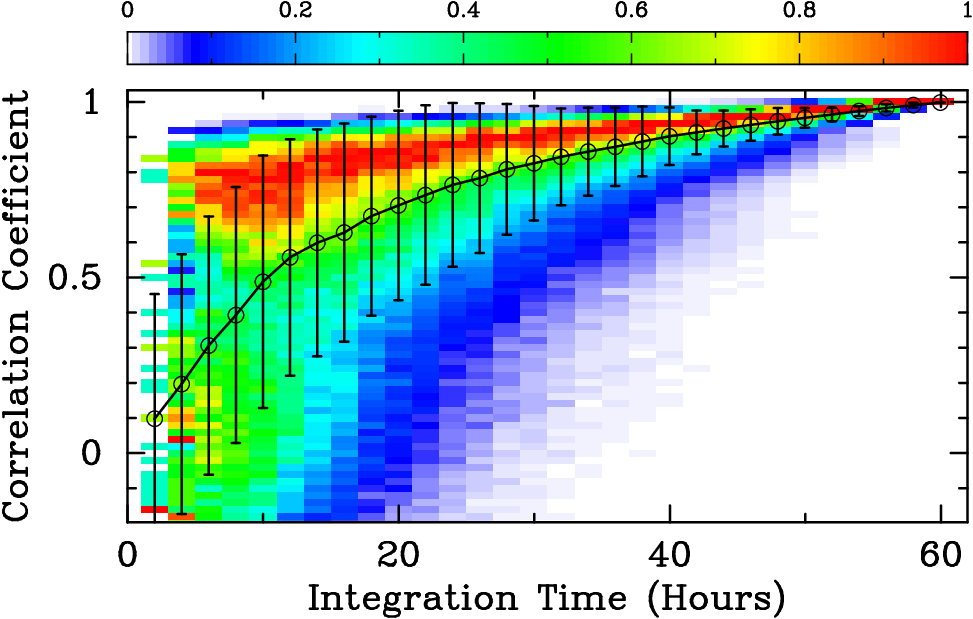}}
\subfigure{\includegraphics[width=0.2\textwidth,angle=-90]{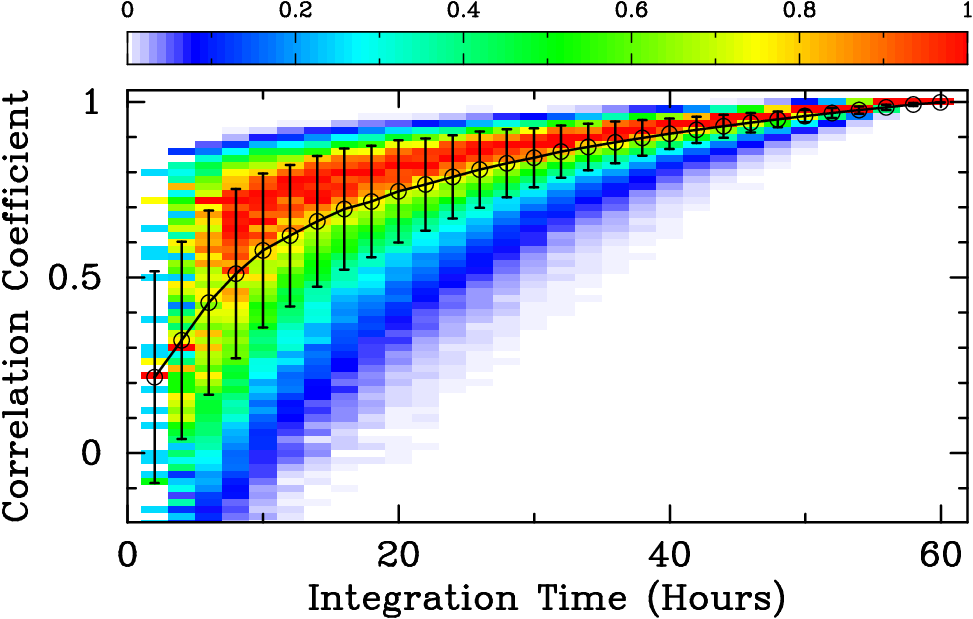}}
\caption{Same as in Figure~\ref{fig_boot} but now for simulated data
(see text). The left, centre and right panels show results when randomly
chosen 2, 4 and 8 of the 30 simulated profiles were modified to obtain
the net average profile shape similar to that in upper panel of
Figure~\ref{fig_mep1}.}
\label{fig_boot2}
\end{figure*}
\subsection{Deep search using March 2014 data}\label{sub_2014}
Out of the 36 observing sessions conducted in 2014, 2 sessions were
unusable due to severe
contamination from RFI. In 4 other sessions, although typically less than
1\% of the samples were identified as RFI-contaminated, data
showed indications of faint RFI and abrupt jumps in power levels.
To minimize effects of such systematics on the search sensitivity, data
from all the above mentioned sessions were excluded from any further processing.
To carry out deep search for a dispersed signal using data from the remaining
30 sessions, we used an up-to-date timing model of J1732$-$3131 obtained
from the gamma-ray data \citep{Kerr15}, to predict the pulsar's rotation
period and phase
at different observing epochs using the pulsar timing software
\textsc{Tempo2}\footnote{For more information about \textsc{Tempo2}, please
refer to the website: \url{http://www.atnf.csiro.au/research/pulsar/tempo2/}.}.
The deep search did not result in any significant signal above our detection
threshold of 8$\sigma$.
\par
To probe a possible underlying periodic signal fainter than our detection
threshold, we dedispersed the data using a DM of 15.5\ppcc\, and computed
average profiles for individual sessions\footnote{The original
detection of the pulsar suggested the DM to be 15.44$\pm$0.32\ppcc\,
\citepalias{MAD12}. Our chosen value of 15.5\ppcc\, is consistent with
this, and choosing the DM to be 15.44\ppcc\, would not have changed
any of the results presented here --- neither qualitatively nor
quantitatively.}. To compute the net average profile, the
individual profiles were weighted and added coherently in the pulsar's
rotation phase. The weights for individual profiles were chosen to be
directly proportional to the product of effective bandwidth and integration
time, i.e., inversely proportional to the expected variance in
the average profile. The 327\,MHz net average profile integrated over an
effective duration of 60\,hours is shown in Figure~\ref{fig_mep1} along
with the average profiles from individual observing sessions. The net
average profile exhibits a peak-to-peak S/N of nearly 6. For comparison,
average profile of the pulsar at 34\,MHz is overlaid on the net average
profile in the upper panel. The two profiles are manually aligned, since
the uncertainty in DM is not adequate enough to phase-align the profiles
at such widely separated frequencies. A striking resemblance between the
two profiles is clearly evident.
Figure~\ref{fig_mep2} shows the dedispersed and averaged
spectrogram of all the data presented in Figure~\ref{fig_mep1}, and
demonstrates that the faint periodic signal is uniformly present across
the observing bandwidth.
\subsubsection{Shape of the net average profile: a faint underlying
signal or artifacts ?}
Before we quantify the similarity between the average profiles at 327 and
34\,MHz, it is important to assess whether the average profile shape has
uniform contribution from all the sessions or dominated by artifacts
in a single or a few individual profiles. A conventional way to test
this is to examine the significance (e.g., peak S/N or reduced chi-square)
of the average profile as data from successive sessions are integrated.
Given the low S/N of the net average profile, we have used a bootstrap
method to
probe whether just a few sessions are affecting the average
profile shape. For this purpose, we examine a correlation-based
figure-of-merit (FoM) as a function of the integration time.
Since the integration time for each of the individual
average profiles is 2\,hours, we have
chosen the step between consecutive trial integration times in our bootstrap
method also to be 2\,hours. For each of the trial integration time, we
randomly choose a sample of appropriate number of individual profiles,
compute a partial average profile using this sample, and measure its
cross-correlation coefficient with the net average profile as a
FoM. Each bootstrap sample corresponding to a trial integration
time consists of choosing $10^4$ non-redundant sample combinations of
sessions using a reservoir sampling algorithm \citep{Jeff85}, and computing
the FoM for each of the combinations. This approach allows
us to obtain a distribution of figures of merit for each of the trial
integration time, with the median of the distribution assigned as the
average FoM. For trial integration
times involving less than 4 or more than 26 sessions, all the possible
combinations of individual profiles are used to compute the FoM distribution.
\par
The average FoM estimated from the above bootstrapping is shown in
Figure~\ref{fig_boot} as a function of integration time, along with the
colour-coded maps of corresponding distributions (histograms) with their peaks
normalized to 1. The uniform distributions and smooth monotonic increase
in the average FoM
with the integration time is clearly evident, and strongly suggests that
the shape of the net average profile shown in Figure~\ref{fig_mep1} represents
a faint signal consistently present in individual profiles obtained from
different sessions.
\par
To demonstrate how presence of artifacts in a few individual
profiles would have reflected in the above analysis, we simulated 30
normally distributed noise profiles. The root-mean-square (rms) of the
noise in these profiles was kept same as that of the observed profiles.
Then a predecided number of profiles were modified in such a way that
the average of all 30 profiles becomes similar to the net average profile
in Figure~\ref{fig_mep1}. For this purpose, intensity of a scaled-version
of the net average profile was randomly distributed between the selected
profiles. The profiles to be modified were themselves chosen randomly.
Figure~\ref{fig_boot2} shows results of the bootstrapping analysis for
three different cases when 2, 4 and 8 profiles were modified as mentioned
above. The first two cases, i.e., when just 2 and 4 profiles are responsible
for the average profile shape, result in glaringly different profile
significance distributions. The third case starts approaching the uniform
distributions and smooth increase in the average FoM for real data shown
in Figure~\ref{fig_boot}. However, minor differences are still noticeable,
e.g., the average FoM rises much more sharply till integration times of
about 20\,hours. The results become nearly indistinguishable from
Figure~\ref{fig_boot} when 16 profiles are modified (not shown but assessed
separately). Hence, the shape of the net average profile is contributed
by a significant number of sessions, perhaps all, and certainly not only
by just a few.
\begin{figure}
\begin{center}
\includegraphics[width=0.24\textwidth,angle=-90]{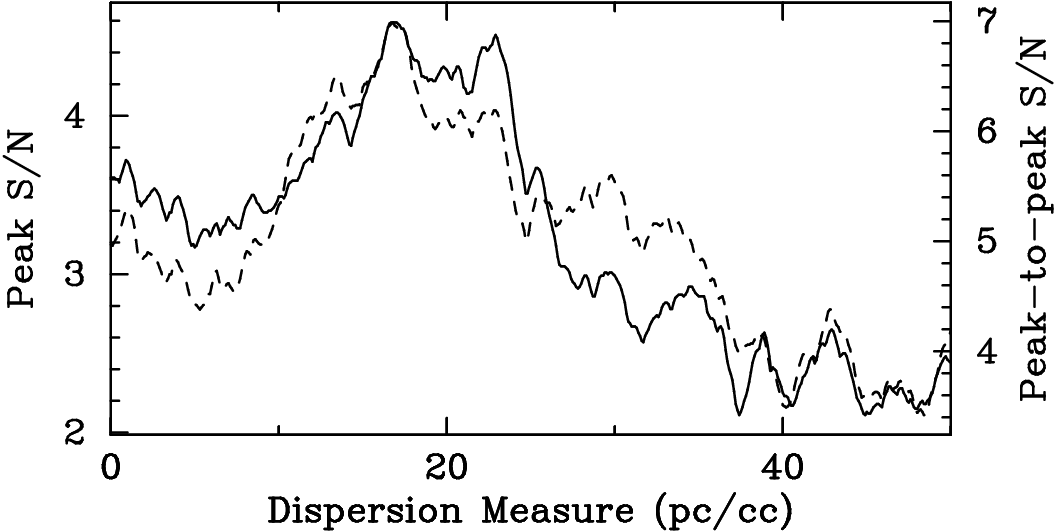}
\caption{The continuous and dashed lines show the peak S/N and the peak-to-peak
S/N of the net average profile as a function of trial DM.}
\label{fig_dmtune}
\end{center}
\end{figure}
\subsubsection{The net average profile: association with the pulsar}
The 327 and 34\,MHz profiles shown in the upper panel of
Figure~\ref{fig_mep1} exhibit striking resemblance, and both are
consistent with each other within the noise uncertainties.
As a quantitative measure of the similarity, the \emph{normalized}
cross-correlation coefficient between the two profiles is estimated
to be 0.72.
\par
To estimate the chance probability of obtaining a net average profile with the
observed peak-to-peak S/N of nearly 6 and exhibiting striking similarity with
the 34\,MHz profile, we performed Monte-Carlo (MC) simulations. Each individual
realization in our MC simulation involves generating 30 random (normal distribution)
noise profiles,
computing an average noise profile from these, and cross-correlating the average
profile with the 34\,MHz profile. To be compatible with the profiles shown
in Figure~\ref{fig_mep1}, each of the random noise profiles is also smoothed
with a $30\degr$\,($\sim0.08$ in normalized
pulse-phase) wide window. The resultant average noise profile is
cross-correlated with the 34\,MHz profile at all possible phase-shifts to
determine the maximum \emph{normalized} correlation coefficient. The maximum
correlation coefficient and the peak-to-peak S/N of the average noise profile
are noted down. From simulations of 1~million ($10^6$) such independent
realizations, only $0.18\%$ times the average noise profile was found to
be having peak-to-peak S/N more than 5.5 as well as the maximum
correlation coefficient $\geq0.72$. Hence, the probability of obtaining
the 327\,MHz profile with its measured significance and similarity to the
34\,MHz profile just by chance is only 0.0018. In other words, the 327\,MHz
profile is consistent with being originated from the same source as the
34\,MHz profile at a confidence level of $99.82\%$.
\par
The confidence gained from the above MC simulations also motivates to
explore the profile significance as a function of DM. Variation in the
profile significance in terms of the peak~S/N as well as peak-to-peak~S/N
with DM, obtained from our deep search methodology mentioned earlier, are
shown in Figure~\ref{fig_dmtune}. The profiles corresponding to the individual
trial DMs were smoothed by a 30\degr\,wide window before computing the
significance. Although the significance is low, peak~S/N as well as
peak-to-peak~S/N are suggestive of a nominal DM of $16\pm7$\ppcc\, which
is consistent with DM estimated from the original detection, and provides
an independent support
for the low-significance net average profile to be originated from the pulsar.
\subsection{Deep search using April 2015 data}
Compared to the 2014 observations, data from the observing sessions conducted
in 2015 showed overall more occurrences of RFIs. Strong RFIs and visible
system artifacts made data from 8 observing sessions unusable, while those
from 3 other sessions showed hints of low-level RFIs, and hence excluded
from further processing.
Deep search for the periodic signal from the pulsar using data from the
remaining 26 sessions, amounting to a total of about 39~hours of integration
time, did not result in any significant candidate above our detection
threshold of $8\sigma$.
\par
Following the procedure detailed in Section~\ref{sub_2014}, we used the
data from above 26 sessions to compute the net average profile dedispersed
at 15.5\ppcc. Note that the integration time for this profile is about
39~hours, as against 60~hours for the profile shown in Figure~\ref{fig_mep1}.
Furthermore, sensitivity of the telescope during these observations was
lower\footnote{A set of bright sources are regularly observed to monitor
the sensitivity of the telescope. The factor mentioned in the main text
is assessed using these observations.} by a factor of 1.2$-$1.3, compared
to that during the 2014 observations. Due to these factors, the peak-to-peak
S/N of the average profile obtained from 2015 observations barely reaches 3.5$-$4.
The average profiles can also be added in-phase to those from 2014 observations
to achieve effectively a larger integration time. However, due to poorer
sensitivity of the telescope during these observations, the effective
increase in the integration time will be about 23--27\,hours, which translates
to only about 1$\sigma$ increase in the profile significance. Even this
feeble enhancement is subjective to quality of the data being added, such
as being free from any underlying faint RFIs. The 2015 data are found
to be of overall poorer quality than the 2014 data. In any case, combining
the observations from the two years did not improve the 2014 average profile
by any significant amount.
\subsection{Search for continuum emission}
Given the large pulse duty cycle, and possible emission at all pulse phases,
the pulsar might be better suited to be detected as a source of continuous
emission using interferometric observations. With this motivation, we used
GMRT archival data at 330\,MHz, from nearly 4\,hours long observations
conducted in March 2004 (proposal code: 05SBA01) towards a nearby source
{G355.5+00}.
The rms noise obtained at the centre of the field is
0.6\,mJy/beam. However, the pulsar is 61$'$ away from the pointing centre,
and the primary beam corrected rms noise at the location of the
pulsar is about 3.4\,mJy/beam. No source could be seen within 5$'$ of the
position of the pulsar (17h32m33.54s, $-$31d31$'$23$''$; J2000). Therefore, we
put a $3\sigma$ continuous emission upper limit of 10\,mJy from the pulsar.
\subsection{Flux density and spectral index estimates}
To estimate the sky background temperature towards the pulsar,
we used a low frequency sky map generating program,
\textsl{LFmap},\footnote{\url{http://www.astro.umd.edu/~emilp/LFmap/}}
to construct a sky map at 326.5\,MHz. LFmap scales the 408 MHz all-sky
map of \citet{Haslam82} to the desired low frequency, taking into account
the CMB, isotropic emission from unresolved extra-galactic sources, and
the anisotropic Galactic emission. By computing a weighted average of
the 326.5\,MHz map at several points across the beam, using the following
theoretical beam-gain pattern:\\
$P(RA,DEC)=\sinc^2{\left(\frac{b\,\sin{(RA)}}{\lambda}\right)}
\times\sinc^2{\left(\frac{a\,\sin{(DEC)}}{\lambda}\right)}$,\\
where $a=\cos{(DEC)}\times530$\,m, $b=30$\,m and $\lambda=0.92$~m,
the sky background temperature towards the pulsar is estimated to be 610\,K.
Assuming an effective bandwidth of 13.6\,MHz (85\% of the total bandwidth),
50\% pulse duty cycle, receiver temperature of 150\,K, and an effective
collecting area of 7500\,m$^2$ (55\% of physical area projected towards the
pulsar), we estimate the pulsar's average flux density to be in the range
0.5$-$0.8\,mJy (for profile-S/N to be in the range 3$-$5 achieved from
60\,hours of integration; Figure~\ref{fig_mep1}).
\par
The above flux density estimate when combined with that
at 34\,MHz \citepalias{MA14} and assuming no turn-over in the spectrum,
suggests the spectral index of the pulsar to be in the range $-2.4$ to
$-3.0$. This range is consistent with the spectral index upper limit
of $-2.3$ suggested in \citetalias{MA14}. We would like to emphasize
that the above spectral index estimate will remain unaffected even if
the actual pulse duty cycle happens to be different from what we have
assumed. This is due to the fact that the profile widths are similar
at 34 and 327\,MHz (see Figure~\ref{fig_mep1}), and same pulse duty
cycle (50\%) has been assumed for estimating the flux densities at both
the frequencies.
\par
The continuum emission upper limit at 330\,MHz presented in the previous
subsection is consistent with the flux density of the pulsar estimated
above. Recently, \citet{Frail16} have reported a 3$\sigma$ flux
density upper limit of 24\,mJy/beam towards this pulsar at 150 MHz.
Assuming no turn-over in the spectrum, the spectral index deduced
above predicts a flux density of 3$-$8\,mJy at 150\,MHz, consistent
with the result from \citet{Frail16}.
\par
The steep spectrum index suggests the pulsar's flux density at 1400\,MHz
to be 6$-$24\,$\mu$Jy, consistent with the upper limit from earlier
searches \citep{Ray11}.
The flux density is also comparable to that of the other two very
faint pulsars mentioned in Section 1 (J0106+4855 and J1907+0602). However,
J1732$-$3131 is relatively nearby implying a lower luminosity. Indeed the
1400\,MHz pseudo-luminosity\footnote{The pseudo-luminosity is defined as
1400\,MHz flux density times the square of the distance, and takes in to
account the uncertainties in the flux density at 327\,MHz and the spectral
index.} of the pulsar is only 2.2$-$8.9\,$\mu$Jy\,kpc$^2$. If we use the
NE2001 electron density model \citep{CL02} to derive distance from DM, then
the above range suggests J1732$-$3131 to be the least luminous among all the
pulsars for which 1400\,MHz flux density estimates are known
\citep[ATNF pulsar catalog;][]{Manchester05}! Even if we use the more recent
electron density model by \citet{YMW16}, only a few pulsars have pseudo-luminosity
less than that of J1732$-$3131. Hence, J1732$-$3131 is one of the least luminous
radio pulsars known to date !
\subsection{Profile morphology and implications to
viewing geometry and emission mechanism}
At this stage, we can perhaps review whether the 34\,MHz detection
of the pulsar was indeed benefitted by its correspondingly larger emission
beam at low frequency. While strong constraints on the viewing geometry
can be obtained only from polarization data, we can seek hints from the
profile morphology and the radio spectrum.
A grazing line of sight would tend to miss the high frequency radio
emission beam and give rise to a steep spectrum
\citep[e.g., B0943+10 has a spectral index of -2.9;][]{MMS00}.
The mean spectral index
for normal pulsars has been estimated to be $-1.4\pm1.0$ \citep{Bates13}.
With the spectral index in the range $-2.4$ to $-3.0$, radio spectrum of
J1732$-$3131 is indeed on the steeper side.
However, more than 10\% of the pulsars with characterized
radio spectra have spectral indices in the range deduced for J1732$-$3131
(ATNF pulsar catalog),
suggesting that the present constraints are too loose to be interpreted
in the present context.
\par
A grazing sight line would also imply a significant evolution of pulse-width
and/or number of pulse components with frequency. The similarity of J1732$-$3131's
pulse profiles at frequencies separated by a factor of nearly 10 (at 34 and
327\,MHz; Figure~\ref{fig_mep1}), suggests that there is no \emph{major}
evolution of the profile in this frequency range. However, minor evolution
that is hindered by very low-S/N in both the profiles can not be excluded.
Hence, the current data do not present a strong support for a favorable
viewing geometry to be responsible for the pulsar's original detection at
low frequency.
\par
The radio and gamma-ray profiles of the pulsar appear to have similar
morphology (see Figure 4 of \citetalias{MAD12}). Profile morphology and
phase-alignment of the radio and gamma-ray profiles could give important
clues about the location of the emission sites. The propagation of radio
signals through the dispersive ISM introduces a delay
$\Delta t \propto {\rm DM}\times\nu^{-2}$. The uncertainty in DM
from the original detection (15.44$\pm$0.32\,\ppcc) was not adequate enough
to probe the phase-alignment of the 34\,MHz profile with its gamma-ray
counterpart. However, at 327\,MHz an uncertainty in DM as large as
0.3\,\ppcc\ translates to a delay equivalent to only about 0.06 of the
pulsar's rotation period, allowing us to examine the phase-alignment
between the radio and the gamma-ray profiles.
\begin{figure}
\begin{center}
\includegraphics[width=0.28\textwidth,angle=-90]{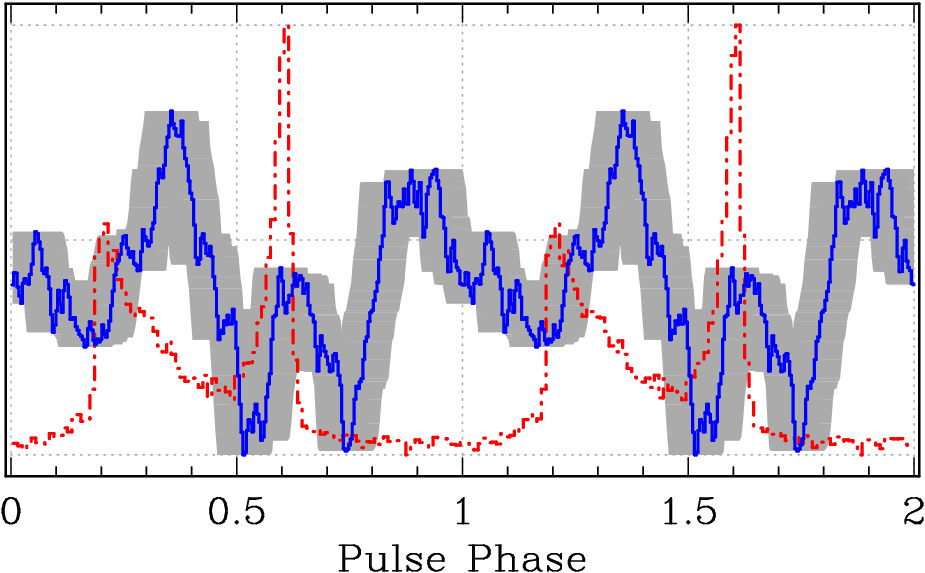}
\caption{Phase-aligned radio and gamma-ray profiles of J1732$-$3131.
The continuous blue line shows the 327\,MHz average profile from
2014 observations and the dashed-dotted-dashed red curve shows the
gamma-ray profile. The horizontal extent of the grey shade on either
side of the
radio profile indicates the uncertainty in phase corresponding
to 1$\sigma$ error in DM. For clarity, both the profiles are duplicated
and plotted over a range spanning two rotations of the pulsar.}
\label{rg_profs}
\end{center}
\end{figure}
\par
The dispersive delay ($\Delta t$) as well as delays associated with the
configuration of the telescope relative to the solar system barycenter,
are accounted for by \textsc{Tempo2} while predicting the pulsar's phase
at a given observing epoch. The instrumental delay associated with the
ORT was estimated by fitting a `jump' in pulse arrival times of a known
fast pulsar obtained at ORT and GMRT using \textsc{Tempo2}. The delay
was confirmed by independent tests, and was taken in to account by
modifying the recorded start time of the individual observations appropriately.
The gamma-ray photon arrival times are
converted to the geocenter to remove the effects of the Fermi LAT spacecraft
motion around the earth. The corrected arrival times are then used with
\textsc{Tempo2} in predictive mode to compute an average profile.
To probe
the phase-alignment of the radio and gamma-ray light curves, we compute
the arrival time correction from the geocenter to the ORT as well as the
associated phase-correction. We use these corrections to convert the gamma-ray
light curve to the ORT, and compare it with the radio profile observed at
the site. We successfully tested our procedure by reproducing
the known phase offset of 0.44 between the radio and gamma-ray profiles
of the pulsar J0437$-$4715 \citep{fermi_catalog13}.
\par
The phase-aligned radio and gamma-ray profiles of J1732$-$3131 are
plotted in Figure~\ref{rg_profs}. The lag ($\delta$) of the main peak
in the gamma-ray profile with respect to the highest peak in the radio
profile is {$0.24\pm0.06$} in pulse-phase\footnote{Difference between
centroids of the two profiles also gives the lag to be consistent with
$0.24\pm0.06$. The small radio peak at pulse-phase of about {0.65} in
Figure~\ref{rg_profs}, which is also noticeably present in the 34\,MHz
profiles \citepalias[see][]{MAD12,MA14}, could be aligned with the main
peak in the gamma-ray profile. However, the S/N of this peak in all the
radio detections so far is too low to claim its possible alignment.}.
Note that the the uncertainty of 0.06
is entirely due to that in DM, and any contribution from the statistical
uncertainty on the position of the peaks is much smaller.
\par
For majority of young gamma-ray pulsars, the phase difference between
the leading and trailing peaks in gamma-ray profile ($\Delta$) has
been found to be anti-correlated with $\delta$. This relationship
between $\Delta$ and $\delta$ was initially shown by \citet{RY95}
to be a general property of the outer-magnetosphere models with
caustic pulses. Majority of the young and middle-aged gamma-ray
pulsars detected by Fermi LAT have confirmed to the $\Delta - \delta$
relationship \citep{fermi_catalog13}. More recently, \citet{Pierbattista16}
have constrained several $\gamma$-ray geometrical models by comparing
simulated and
observed light curve morphological characteristics, including the
correlation of $\delta$ with other observable quantities. Assuming a
vacuum-retarded dipole magnetic field, they show that the Outer
Gap (OG) model \citep{CRZ00} and the One Pole Caustic (OPC) model
\citep[an alternative formulation of the OG;][]{RW10} best explain
the trends between various observed and deduced parameters.
Our deduced value of $\delta=0.24\pm0.06$ suggests that J1732$-$3131
also follows the common trends (like correlation of $\delta$ with
rotation period and $\Delta$) exhibited by a large fraction of
radio-loud gamma-ray pulsars \citep[$\Delta=0.42$, and see, for
example, Figures 12 and 13 in][]{Pierbattista16}.
Hence, following the conclusions of \citet{Pierbattista16}, the observed
radio and gamma-ray light curves of J1732$-$3131 are also best explained
by the OG and OPC models.
\section{Summary}
In the previous sections, we have presented details of our extensive
observations, and deep search for periodic signal from the gamma-ray
pulsar J1732$-$3131 at 327\,MHz using the Ooty radio telescope. Despite
the high background sky temperature, our deep integration allowed us to
probe very faint periodic signal from the pulsar. Using nearly 60\,hours
of observations conducted in March 2014, we have presented detection of
periodic signal at a DM of 15.5\ppcc\, from the pulsar at a confidence
level of 99.82\%. We estimate the 327\,MHz flux density of the pulsar
to be 0.5$-$0.8\,mJy, and the spectral index in the range
from $-2.4$ to $-3.0$.
The 1400\,MHz pseudo-luminosity of the pulsar is only
2.2$-$8.9\,$\mu$Jy\,kpc$^2$, and suggests the pulsar to be one of the
least luminous pulsars known to date!
We also phase-aligned the radio and gamma-ray
profiles, and measured the phase-offset between the main peaks in the
two profiles to be $0.24\pm0.06$. This non-zero phase-lag favors the
models wherein the gamma-ray emission originates in the outer magnetosphere
of the pulsar.
\par
J1732$-$3131 was detected during a rare enhancement in its flux density,
most likely due to scintillation \citepalias{MAD12}.
Subsequent detections of the pulsar using deep follow-up observations
\citepalias[][and this work]{MA14} have been possible only since its
DM was known from the first detection \citepalias{MAD12}. So, it is
possible that some of the radio-quiet gamma-ray pulsars might actually
be very faint radio sources, and hence not detected in the radio searches
using current generation telescopes. The high sensitivity of upcoming radio
telescopes like square kilometre
array (SKA) and the five hundred meter aperture spherical telescope (FAST)
will enable radio detection, and facilitate better studies of such pulsars.
\section*{Acknowledgements}
YM acknowledges use of the funding from the European Research Council under
the European Union's Seventh Framework Programme (FP/2007-2013)/ERC Grant
Agreement no. 617199.
BCJ, PKM and MAK acknowledge support from the Department of Science and
Technology grant DST-SERB Extra-mural grant EMR/2015/000515.
BCJ, PKM, MAK and AKN acknowledge support from TIFR XII plan grants 12P0714
and 12P0716.
YM would like to thank Cees Bassa, Benjamin Stappers and Andrew Lyne for
providing profiles of a few normal pulsars that were useful in cross-checking
the instrumental delays.
YM, MAK, AKN, BCJ and PKM acknowledge the kind help and support provided
by the members of the
Radio Astronomy Centre, Ooty, during these observations. ORT is operated
and maintained at the Radio Astronomy Centre by the National Centre for
Radio Astrophysics. We have used the archival data provided by the GMRT.
GMRT is run by the National Centre for Radio Astrophysics of the Tata
Institute of Fundamental Research.

\bsp	
\label{lastpage}

\begin{thebibliography}{}
\makeatletter
\relax
\def\mn@urlcharsother{\let\do\@makeother \do\$\do\&\do\#\do\^\do\_\do\%\do\~}
\def\mn@doi{\begingroup\mn@urlcharsother \@ifnextchar [ {\mn@doi@}
  {\mn@doi@[]}}
\def\mn@doi@[#1]#2{\def\@tempa{#1}\ifx\@tempa\@empty \href
  {http://dx.doi.org/#2} {doi:#2}\else \href {http://dx.doi.org/#2} {#1}\fi
  \endgroup}
\def\mn@eprint#1#2{\mn@eprint@#1:#2::\@nil}
\def\mn@eprint@arXiv#1{\href {http://arxiv.org/abs/#1} {{\tt arXiv:#1}}}
\def\mn@eprint@dblp#1{\href {http://dblp.uni-trier.de/rec/bibtex/#1.xml}
  {dblp:#1}}
\def\mn@eprint@#1:#2:#3:#4\@nil{\def\@tempa {#1}\def\@tempb {#2}\def\@tempc
  {#3}\ifx \@tempc \@empty \let \@tempc \@tempb \let \@tempb \@tempa \fi \ifx
  \@tempb \@empty \def\@tempb {arXiv}\fi \@ifundefined
  {mn@eprint@\@tempb}{\@tempb:\@tempc}{\expandafter \expandafter \csname
  mn@eprint@\@tempb\endcsname \expandafter{\@tempc}}}

\bibitem[\protect\citeauthoryear{{Abdo} et~al.,}{{Abdo} et~al.}{2010}]{Abdo10}
{Abdo} A.~A.,  et~al., 2010, ApJ, 711, 64

\bibitem[\protect\citeauthoryear{{Abdo} et~al.,}{{Abdo}
  et~al.}{2013}]{fermi_catalog13}
{Abdo} A.~A.,  et~al., 2013, ApJS, 208, 17

\bibitem[\protect\citeauthoryear{{Bates}, {Lorimer}  \& {Verbiest}}{{Bates}
  et~al.}{2013}]{Bates13}
{Bates} S.~D.,  {Lorimer} D.~R.,   {Verbiest} J.~P.~W.,  2013, MNRAS, 431, 1352

\bibitem[\protect\citeauthoryear{{Brazier} \& {Johnston}}{{Brazier} \&
  {Johnston}}{1999}]{BJ99}
{Brazier} K.~T.~S.,  {Johnston} S.,  1999, MNRAS, 305, 671

\bibitem[\protect\citeauthoryear{{Caraveo}}{{Caraveo}}{2014}]{Caraveo14}
{Caraveo} P.~A.,  2014, ARA\&A, 52, 211

\bibitem[\protect\citeauthoryear{{Cheng}, {Ruderman}  \& {Zhang}}{{Cheng}
  et~al.}{2000}]{CRZ00}
{Cheng} K.~S.,  {Ruderman} M.,   {Zhang} L.,  2000, ApJ, 537, 964

\bibitem[\protect\citeauthoryear{{Cordes}}{{Cordes}}{1978}]{Cordes78}
{Cordes} J.~M.,  1978, ApJ, 222, 1006

\bibitem[Cordes \& Lazio(2002)]{CL02}
Cordes, J.~M., \& Lazio, T.~J.~W.\ 2002, arXiv:astro-ph/0207156 

\bibitem[\protect\citeauthoryear{{Frail}, {Jagannathan}, {Mooley}  \&
  {Intema}}{{Frail} et~al.}{2016}]{Frail16}
{Frail} D.~A.,  {Jagannathan} P.,  {Mooley} K.~P.,   {Intema} H.~T.,  2016,
  ApJ, 829, 119

\bibitem[\protect\citeauthoryear{{Haslam}, {Salter}, {Stoffel}  \&
  {Wilson}}{{Haslam} et~al.}{1982}]{Haslam82}
{Haslam} C.~G.~T.,  {Salter} C.~J.,  {Stoffel} H.,   {Wilson} W.~E.,  1982,
  A\&AS, 47, 1

\bibitem[\protect\citeauthoryear{{Jeffrey}}{{Jeffrey}}{1985}]{Jeff85}
{Jeffrey} S.~V.,  1985, ACM Transactions on Mathematical Software (TOMS), 1, 37

\bibitem[\protect\citeauthoryear{{Kerr}, {Ray}, {Johnston}, {Shannon}  \&
  {Camilo}}{{Kerr} et~al.}{2015}]{Kerr15}
{Kerr} M.,  {Ray} P.~S.,  {Johnston} S.,  {Shannon} R.~M.,   {Camilo} F.,
  2015, ApJ, 814, 128

\bibitem[\protect\citeauthoryear{{Maan}}{{Maan}}{2014}]{ythesis}
{Maan} Y.,  2014, PhD thesis, Indian Institute of Science, Bangalore, India

\bibitem[\protect\citeauthoryear{{Maan}}{{Maan}}{2015}]{Maan15}
{Maan} Y.,  2015, ApJ, 815, 126

\bibitem[\protect\citeauthoryear{{Maan} \& {Aswathappa}}{{Maan} \&
  {Aswathappa}}{2014}]{MA14}
{Maan} Y.,  {Aswathappa} H.~A.,  2014, MNRAS, 445, 3221

\bibitem[\protect\citeauthoryear{{Maan}, {Aswathappa}  \& {Deshpande}}{{Maan}
  et~al.}{2012}]{MAD12}
{Maan} Y.,  {Aswathappa} H.~A.,   {Deshpande} A.~A.,  2012, MNRAS, 425, 2

\bibitem[\protect\citeauthoryear{{Malofeev}, {Malov}  \&
  {Shchegoleva}}{{Malofeev} et~al.}{2000}]{MMS00}
{Malofeev} V.~M.,  {Malov} O.~I.,   {Shchegoleva} N.~V.,  2000, Astron. Rep.,
  44, 436

\bibitem[\protect\citeauthoryear{{Manchester}, {Hobbs}, {Teoh}  \&
  {Hobbs}}{{Manchester} et~al.}{2005}]{Manchester05}
{Manchester} R.~N.,  {Hobbs} G.~B.,  {Teoh} A.,   {Hobbs} M.,  2005, AJ, 129,
  1993

\bibitem[\protect\citeauthoryear{{Naidu}, {Joshi}, {Manoharan}  \&
  {Krishnakumar}}{{Naidu} et~al.}{2015}]{Naidu15}
{Naidu} A.,  {Joshi} B.~C.,  {Manoharan} P.~K.,   {Krishnakumar} M.~A.,  2015,
  ExA, 39, 319

\bibitem[\protect\citeauthoryear{{Pierbattista}, {Harding}, {Gonthier}  \&
  {Grenier}}{{Pierbattista} et~al.}{2016}]{Pierbattista16}
{Pierbattista} M.,  {Harding} A.~K.,  {Gonthier} P.~L.,   {Grenier} I.~A.,
  2016, A\&A, 588, A137

\bibitem[\protect\citeauthoryear{{Pletsch} et~al.,}{{Pletsch}
  et~al.}{2012}]{Pletsch12a}
{Pletsch} H.~J.,  et~al., 2012, ApJ, 744, 105

\bibitem[\protect\citeauthoryear{{Ray} et~al.,}{{Ray} et~al.}{2011}]{Ray11}
{Ray} P.~S.,  et~al., 2011, ApJS, 194, 17

\bibitem[\protect\citeauthoryear{{Romani} \& {Watters}}{{Romani} \&
  {Watters}}{2010}]{RW10}
{Romani} R.~W.,  {Watters} K.~P.,  2010, ApJ, 714, 810

\bibitem[\protect\citeauthoryear{{Romani} \& {Yadigaroglu}}{{Romani} \&
  {Yadigaroglu}}{1995}]{RY95}
{Romani} R.~W.,  {Yadigaroglu} I.-A.,  1995, ApJ, 438, 314

\bibitem[\protect\citeauthoryear{{Swarup} et~al.,}{{Swarup}
  et~al.}{1971}]{Swarup71}
{Swarup} G.,  et~al., 1971, NPhS, 230, 185

\bibitem[\protect\citeauthoryear{{Watters} \& {Romani}}{{Watters} \&
  {Romani}}{2011}]{WR11}
{Watters} K.~P.,  {Romani} R.~W.,  2011, ApJ, 727, 123

\bibitem[Yao et al.(2017)]{YMW16}
Yao, J.~M., Manchester, R.~N., \& Wang, N.\ 2017, \apj, 835, 29

\makeatother
\end{thebibliography}
\end{document}